# Adsorption of silicon on Au(110): an ordered two dimensional surface alloy

Hanna Enriquez[1*], Andrew Mayne[1], Abdelkader Kara[2], Sébastien Vizzini[3], Silvan Roth[4], Boubekeur Lalmi[5], Ari P Seitsonen[6], Bernard Aufray[7], Thomas Greber[4], Rachid Belkhou[5], Gérald Dujardin[1] and Hamid Oughaddou[1,8,*]

[1]*Institut des Sciences Moléculaires d'Orsay, ISMO-CNRS, Bât. 210, Université Paris-Sud, F-91405 Orsay, France*
[2]*Department of Physics, University of Central Florida, Orlando, FL 32816, USA*
[3]*CNRS, IM2NP, Université Aix-Marseille, Faculté des Sciences et Techniques de Jérôme, F-13397 Marseille, France*
[4]*Physik Institut der Universität Zürich, Winterthurerstrasse 190, CH-8057 Zurich, Switzerland*
[5]*Synchrotron SOLEIL, F-91192 Gif Sur Yvette, France*
[6]*Physikalisch-Chemisches Institut der Universität Zürich, Winterthurerstrasse 190, CH-8057 Zürich, Switzerland*
[7] *CNRS, CINaM-UPR3118, Campus de Luminy, Marseille F-13288 Cedex 09, France*
[8]*Département de Physique, Université de Cergy-Pontoise, F-95031 Cergy-Pontoise Cedex, France*

*Corresponding authors: Prof. H. Oughaddou and Dr. H. Enriquez

Abstract:

We report on experimental evidence for the formation of a two dimensional Si/Au(110) surface alloy. In this study, we have used a combination of scanning tunneling microscopy, low energy electron diffraction, Auger electron spectroscopy and *ab initio* calculations based on density functional theory. A highly ordered and stable Si-Au surface alloy is observed subsequent to growth of a sub-monolayer of silicon on an Au(110) substrate kept above the eutectic temperature.

Although widely used in electronic device and nanowire technology, the eutectic of the Au-Si binary phase diagram is not yet completely understood. Over the last few decades, the non-equilibrium phases of this system have been investigated; for example, an amorphous Au-Si alloy was produced by splat quenching, providing the first observed metallic glass phase [1]. Metastable bulk crystalline phases were then obtained through different methods and exhibit a wide variety of stoichiometrices (10 to 30% at. Si), and crystal structures [2-6].

More recently, a stable two-dimensional (2D) gold silicide has been evidenced following surface crystallisation of the eutectic $Au_{82}Si_{18}$ liquid above the eutectic temperature $T_E$ = 359°C [7]. It was shown to have an $Au_4Si_8$ composition and a rectangular crystal structure, stable up to 371°C [8]. This 2D crystalline silicide phase has also been obtained under ultra high vacuum (UHV) conditions following deposition of a 3 nm Au layer on a Si(100) substrate and subsequent annealing above the eutectic temperature ($T_E$) [9]. Other Au/Si alloys were also obtained in UHV conditions either at room temperature (RT) [10-12] or by annealing thicker Au films (10-100 nm) deposited on a Si substrate [4,5]. At RT, an Au-Si surface alloy was obtained with an estimated stoichiometry of $Au_3Si$ or $Au_3Si_2$ [10-12] but no evidence of any crystalline structure was found, whereas annealing the thick Au films (10-100 nm) produced crystalline bulk phases with a thickness in the range of 0.2 to 0.9 nm [5].

In general, the growth of Au on silicon substrates has been extensively studied, however, only very few investigations have been performed on the reverse system (silicon on Au substrates) [13,14]. Indeed, the reversal of the deposition sequence Semiconductor/Metal instead of Metal/Semiconductor can have a significant influence on the interface [15], as in the case of Si and Ge deposited on silver substrates [16-36] where new structures have been observed such as Si and Ge tetramers [16-19, 22-24, 36] or silicene nano-ribbons and sheets [25-36].

In this letter, we report our results on the formation of a 2D Si-Au surface alloy obtained by growth, under UHV conditions, of a sub-monolayer coverage of silicon on a bare Au(110) substrate kept at a temperature above the eutectic temperature (between 360-500°C). A 2D surface alloy is formed and is stable at RT, and presents two mirror-symmetric domains with respect to the high symmetry axes of the substrate.

The apparatus in which the experiments were performed are equipped for surface preparation and characterization: an ion gun for surface cleaning, a low energy electron diffractometer (LEED) for structural characterisation, a RT scanning tunnelling microscope (STM) for surface characterization at the atomic scale, and an Auger electron spectrometer (AES) for chemical surface analysis and the calibration of the silicon coverage. The experiments were performed on the same crystal in two different chambers in ISMO-Orsay [37] and at the University of Zurich [38].

The Au(110) sample was cleaned with several sputtering cycles (600 eV $Ar^+$ ions, P = 5 x $10^{-5}$ Torr) and annealing at 450°C until a sharp p(2x1) LEED pattern, reminiscent for the Au(110) missing row reconstruction, was obtained. Silicon, evaporated by direct current heating of a piece of Si wafer, was deposited onto the Au(110) surface. held at temperatures above the eutectic temperature.

Figure 1a shows the p(2x1) LEED pattern characteristic of the bare Au(110) surface reconstruction. During silicon deposition on the Au(110) substrate at 400°C, the 2x1 starts to disappear while a new superstructure appears, becoming sharp at a Si coverage of ~ 0.2 monolayer (ML). The LEED pattern corresponding to this new superstructure (Figure 1b) shows two symmetrical domains with respect to the [001]* and [$\bar{1}$10]* directions of the substrate and presents a two-fold symmetry. The unit cells corresponding to the substrate and to the two domains are indicated. The real-space vectors of the two domains can be extracted from the experimental LEED pattern in the form of two matrices: $\begin{vmatrix} 10 & -1 \\ -2 & 4 \end{vmatrix}$, and $\begin{vmatrix} 10 & 1 \\ 2 & 4 \end{vmatrix}$. The agreement between the observed LEED pattern in Figure 1b and the simulated one in Figure 1c using the reciprocal vectors corresponding to these matrices is remarkably good.

Figure 2 displays an atomically-resolved filled-state STM image of the bare Au(110) surface. The missing-row structure of the Au(110)-(2x1) reconstruction is clearly observed., matching the LEED pattern in Figure 1a. Figure 3 displays an atomically resolved filled-state STM image of the first stage of Si growth on Au(110) equivalent to a Si coverage < 0.1 ML. We observe a coexistence of the bare Au(110)-2x1 surface and a superstructure composed of two domains (1) and (2). The oblique unit cells corresponding to the matrices $\begin{vmatrix} 10 & -1 \\ -2 & 4 \end{vmatrix}$, and $\begin{vmatrix} 10 & 1 \\ 2 & 4 \end{vmatrix}$, respectively, are marked, showing their orientations with respect to the $[\bar{1}10]$ direction. In both domains, the x2 periodicity of the bare Au(110) disappeared, suggesting that Si adsorption is driven by a strong interaction between the silicon and gold atoms.

Figure 4a shows an atomically resolved filled-state STM image recorded at ~ 0.2 Si ML showing one domain with the $\begin{vmatrix} 10 & -1 \\ -2 & 4 \end{vmatrix}$ superstructure. The oblique unit cell of this superstructure is indicated. The STM topography reveals the atomic-scale pattern of the superstructure with the same two-fold symmetry as observed in the LEED pattern, and indicated by the unit cell in Figure 4a. Annealing this structure up to 500°C neither changes the LEED pattern nor the STM images of the superstructure indicating a high thermal stability. This also supports the idea of a strong interaction between Au and Si atoms.

Figure 4b shows a line profile along the line "A" drawn on Figure 4a. The very small z-corrugation (~ 0.04 nm) suggests that the superstructure contains only weak roughness. Therefore, we can assume that all atoms in the unit cell are in the same average plane. The STM images recorded at negative and positive sample biases display the same contrast, which probably implies that we observe the atomic geometry rather than any electronic effect. On this basis, assuming that each spot corresponds to one atom, the pattern in the unit cell can be described by four similar entities. Each entity is composed of three atoms as highlighted in Figure 4. Furthermore, the observed difference in contrast in the STM image indicates one bright atom and two less bright atoms per entity.

In order to propose an atomistic model for the Si adsorption structure starting from the LEED periodicity, the atomically resolved STM topography and the AES calibration (~0.2 ML Si), we performed density functional theory (DFT) [39] calculations. We systematically studied plausible arrangements of Si atoms on Au(110) in the surface unit cell determined above. Since a single layer of Au(110) in that unit cell contains 38 gold atoms, we chose to investigate configurations containing 8-12 Si atoms. We used the VASP code [40] for the calculations with the projector augmented wave method [42], and expanded the orbitals in plane waves up to an energy cut-off of 245 eV. We employed the Perdew-Burke-Ernzerhof (PBE) generalized gradient approximation [41] as the exchange-correlation term.

As stated above, the LEED and AES measurements at the coverage ~0.2 ML of Si on the Au(110) surface no longer show the 2x1 reconstruction. Therefore we used a non-reconstructed layer at the surface of the unit cell. The slab geometry consisted of five layers of substrate, each containing 38 Au atoms, and two lowest layers were kept fixed. We used a DFT-PBE lattice constant of 4.17 Å for Au and sampled the surface Brillouin zone using a 2x4 mesh of k points. In the simulations of STM images we applied the Tersoff-Hamann method with an s-like tip [43].

We first explored several configurations having the 2-fold symmetry observed in the experiments by adsorbing an even number of silicon atoms without removing any Au atoms. Since the number of similar entities within the unit cell is a multiple of 4, we adsorbed 8 and 12 silicon atoms at different sites (hollow, short-bridge, long bridge and top) into the large unit cell on the Au(110) surface in agreement with the AES calibration. After relaxation, each configuration ended up with every silicon atom adsorbed on a hollow site. The contrast in the calculated STM images for all these configurations was not in agreement with the one observed in the experimental images. In other words, Si atoms adsorbed only on 4-fold sites do not introduce the necessary corrugation in the electronic density to reproduce the experimental STM images. Note that during Si adsorption, the Au surface is at high temperature, so diffusion of silicon atoms into the subsurface region may occur. With this in mind, we considered configurations including Si atoms located below the topmost gold layer. In Figure 5 we propose a configuration where 8 silicon atoms occupy hollow sites and 4 silicon atoms are sitting just below 4 gold atoms. After an initial relaxation of the system, the "elevated" Au atoms are shifted slightly from their ideal position on top of the silicon atoms and are located 0.1 nm above the other gold surface atoms, producing the corrugation necessary to explain the bright spots in the entities observed in the experimental STM image (Figure 4), the less bright spots being assigned to the Si atoms in the hollow sites. In Figure 6 we present a simulated STM image of the configuration shown in Figure 5. To make the comparison, the calculated STM image has the same size and bias voltage as the experimental one in Figure 4. The protrusions form the same entities in both the experimental and simulated images, indicating that the model reproduces the experimental image very well.

We stress that we have tested several atomic configurations and only the model proposed in Figure 5 gives a good agreement between the experimental and simulated STM images. We later noticed that this model is metastable and after a full relaxation, the elevated gold atoms move toward the nearest 4-fold site. We believe that in the real system, the surface stress, which is not taken into account in our calculation because of the small size of the unit cell, could stabilize the proposed structure.

The results show strong interactions between Si and Au atoms. Indeed, the Si atoms within the unit cell are bound only to Au. This is a sign that an ordered surface alloy is formed; the alloy has a stoichiometry of $Au_{38}Si_{12}$.

The role of the annealing above the eutectic temperature in the formation of an ordered alloy needs to be underlined. Indeed, alloy formation has been reported for a few ML of Au deposited at RT on Si (111) with stoichiometry of $Au_3Si$ or $Au_3Si_2$, but no order was observed [10]. Another study based on a grazing incident diffraction (GIX) showed that annealing above the eutectic temperature was mandatory to form an ordered alloy structure after the deposition of Au on Si(100) at RT [9]. In our experiment we show that the eutectic temperature must be also reached in the case of Si/Au(110) to obtain an ordered surface alloy.

In conclusion, deposition of a sub-monolayer of Si on Au(110)-(2x1) removes the (2x1) periodicity of the bare Au substrate indicating a phase transition from a 1D missing-row structure to a 2D structure. We have proposed a model for this 2D structure with a stoichiometry close to $Au_3Si$. We have shown that this 2D structure is very stable up to temperatures of 500°C, indicating a strong interaction between silicon and gold atoms, and the formation of an ordered surface alloy.

Acknowledgements:

This work was partially supported by the project SILICENE of the Triangle de la Physique program (AO 2010-2)

**Figure Captions:**

**Figure 1:** a) LEED pattern corresponding to the (2x1) reconstruction of the bare Au(110) surface ($E_p$=55eV), with the unit cell marked. b) LEED pattern corresponding to the superstructure obtained after the deposition of ~0.2 Si ML on Au(110) kept at 400°C. Unit cell vectors corresponding to two mirror-symmetric domains of the surface alloy are indicated by the blue and red arrows and the rectangular p(1x1) unit cell of the Au(110) substrate is drawn in black ($E_p$=55eV). c) A simulation of the LEED pattern corresponding to the superstructure with two mirror-symmetric domains. Again, the unit cell vectors corresponding to two mirror-symmetric domains of the surface alloy are indicated by the blue and red arrows and the rectangular p(1x1) unit cell of the Au(110) substrate is drawn in black.

**Figure 2:** Atomically resolved filled-states STM image corresponding to the bare Au (110) showing the Au(110)-2x1 reconstruction (V = -80 mV, I = 2.2 nA), The unit cell is indicated with a rectangle.

**Figure 3:** Atomically resolved filled-states STM image recorded after the first stage of Si adsorption (<0.1 Si ML), (V = -1.5V, I = 1.7nA). The unit cells of the two mirror-symmetric domains are indicated in blue lines.

**Figure 4:** a) Atomically resolved filled-states STM image recorded at ~ 0.2 Si ML showing a $\begin{vmatrix} 10 & -1 \\ -2 & 4 \end{vmatrix}$ superstructure (V = -1.4V, I = 2.3nA). The unit cell and the two-fold symmetry are indicated. b) Line profiles along the line "A". An entity made of three atoms is indicated by the ellipse.

**Figure 5:** Proposed model where 8 silicon atoms occupy 4-fold sites and 4 silicon atoms are sitting just below 4 gold atoms. a) top view, b) perspective view. The Au atoms of the 1$^{st}$ and the 2$^{nd}$ monolayers are colored in gold and grey respectively. Si atoms are in blue. The Au atoms located on top of Si atoms are colored in pink. The unit cell is indicated with black lines.

**Figure 6:** Simulated STM image corresponding to the model shown in Figure 5. The unit cell is indicated with white lines. The large/strong protrusions are emphasized with circles.

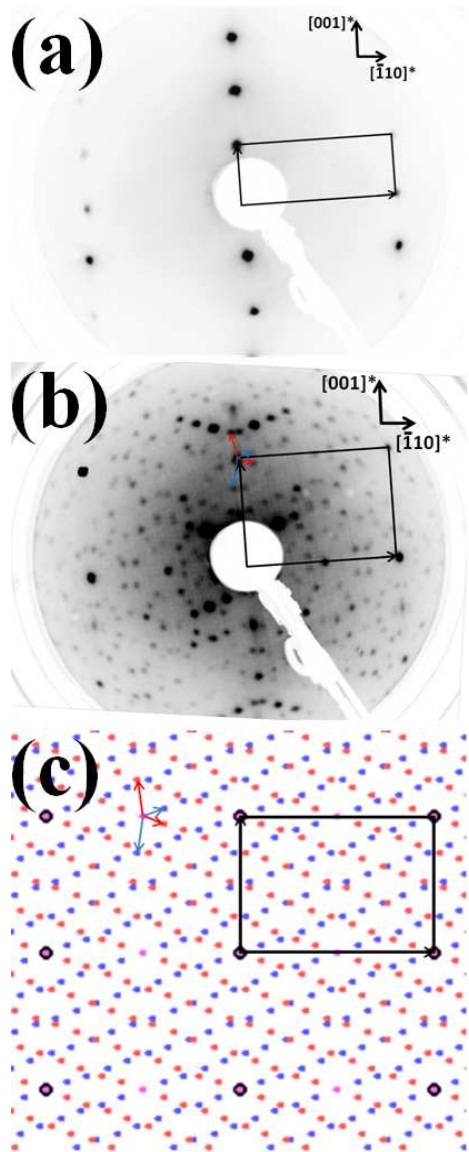

**Figure 1**

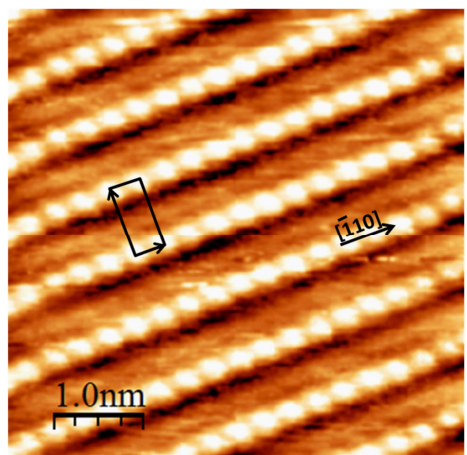

**Figure 2**

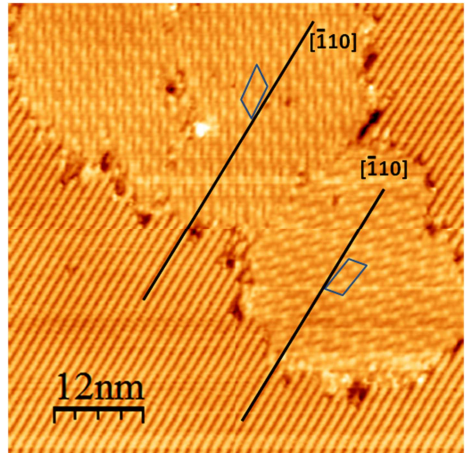

**Figure 3**

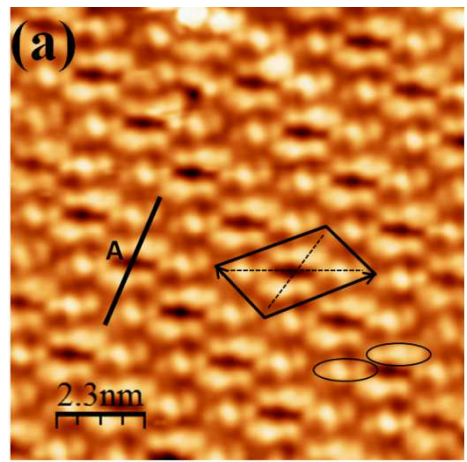

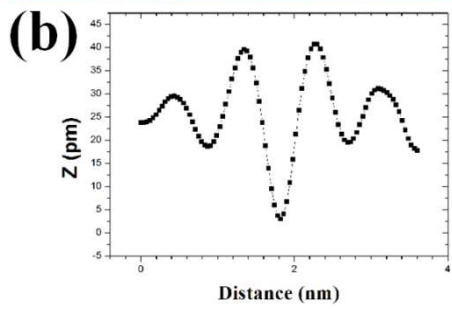

**Figure 4**

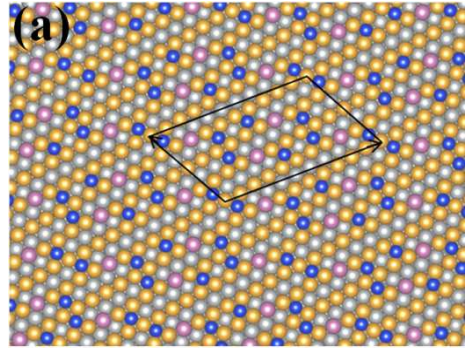

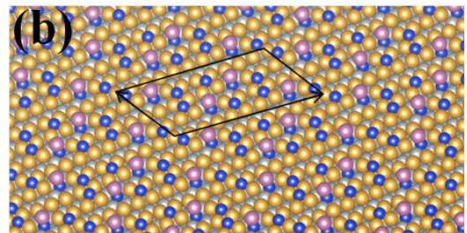

**Figure 5**

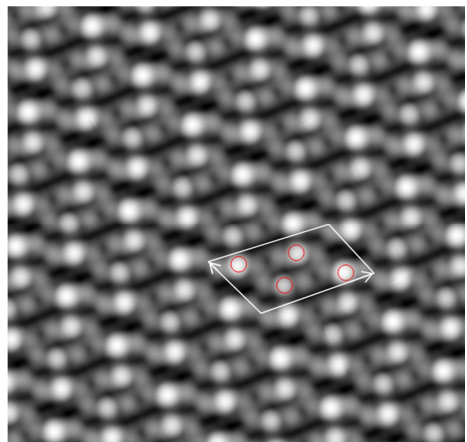

**Figure 6**